\documentclass[a4paper]{revtex4}
\usepackage{graphicx}
\usepackage{fancyhdr}
\usepackage{amsmath}
\newcommand{\be}{\begin{equation}}
\newcommand{\ee}{\end{equation}}
\newcommand{\ba}{\begin{align}}
\newcommand{\ea}{\end{align}}
\newcommand{\bea}{\begin{eqnarray}}
\newcommand{\eea}{\end{eqnarray}}

\usepackage{color}

\begin{document}

\title{One-loop inert and pseudo-inert minima}

%

\author{P.M.~Ferreira}
    \email[E-mail: ]{ferreira@cii.fc.ul.pt}
\affiliation{ISEL - Instituto Superior de Engenharia de Lisboa, Instituto Polit\'ecnico de Lisboa, Portugal}
\affiliation{Centro de F\'{\i}sica Te\'{o}rica e Computacional,
    Faculdade de Ci\^{e}ncias,
    Universidade de Lisboa, Portugal}

\author{B. \'{S}wie{\.z}ewska}
    \email[E-mail: ]{bogumila.swiezewska@fuw.edu.pl}
\affiliation{Faculty of Physics, University of Warsaw, Pasteura 5, 02-093 Warsaw, Poland}

\begin{abstract}
We analyse the differences between inert and pseudo-inert vacua in the 2HDM, both at tree-level
and at one-loop. The validity of tree-level formulae for the relative depth of the potential
at both minima is studied. The one-loop analysis shows both minima can coexist in regions
of parameter space forbidden at tree-level.
\end{abstract}

\maketitle

\thispagestyle{fancy}


\section{Introduction}

The two Higgs doublet model (2HDM)~\cite{Lee:1973iz} is one of the simplest extensions of the Standard
Model (SM) of particle physics, in which an extra scalar doublet is added to the theory.
This  addition originates a richer scalar spectrum than the SM's, and in the versions of the model
wherein CP symmetry is conserved, this includes two CP-even scalars (the lightest $h$ and the heaviest $H$),
a pseudoscalar, $A$, and a charged scalar, $H^\pm$.
For a recent 2HDM review, see~\cite{Branco:2011iw}. The recent discovery of the Higgs boson~\cite{:2012gk,:2012gu}
constrained the 2HDM parameter space, and it has been verified that the model
survives comparison with data. In fact, the 2HDM does a very good job describing the LHC
results~\cite{Chen:2013kt,Belanger:2012gc,Chang:2012ve,Ferreira:2011aa,Chowdhury:2015yja}.
The 2HDM has an interesting phenomenology, including possible spontaneous CP violation,
tree-level scalar-mediated flavour changing neutral currents and, in certain versions of the model, dark matter candidates.

Those versions of the 2HDM correspond to a theory wherein one has imposed a discrete $Z_2$ symmetry on the potential,
the so-called {\em Inert Doublet Model} (IDM)~\cite{Ma:1978,Cao:2007,Barbieri:2006}. Under such a symmetry,
one of the doublets ($\Phi_1$, for instance)
is left unchanged, but the other ($\Phi_2$) is transformed, such that $\Phi_2 \rightarrow -\Phi_2$.
Then, in the IDM, the electroweak symmetry is broken
by a vacuum which preserves the discrete $Z_2$, and as such there is a new quantum number which
must be preserved in all interactions. By choosing the $Z_2$ parity of the particles
of the model, it is simple to ensure that several of the scalars do not couple to fermions at tree-level
(nor indeed possess any triple couplings to gauge bosons). This leads to the lightest
neutral scalar which is odd under $Z_2$ being stable, and as such a prime candidate for dark matter. For works
on the IDM, see for instance~\cite{LopezHonorez:2006,Dolle:2009,Honorez:2010,Sokolowska:2011,Gustafsson:2012,
Swiezewska:2012,Krawczyk:2013jhep,Arhrib:2013,Goudelis:2013,Chakrabarty:2015,Ginzburg:2010,
Sokolowska:2011t,Kanemura:2004,Gil:2012,Cline:2013,Chowdhury:2011,Ma:2006neutrino,Gustafsson:2012neutrino,
Davoudiasl:2014pya}.

The $Z_2$-symmetric 2HDM scalar potential, furthermore, has an interesting vacuum structure. Several types
of extrema are possible already at tree-level, and under specific circumstances, minima of different depths,
leading to different types of physics, may coexist in the
model~\cite{Ivanov:2006yq,Ivanov:2007de,Barroso:2007rr,Ginzburg:2010}. In the current work, we will review the
tree-level analysis pertaining to the coexistence of such minima, and study the impact that one-loop contributions
to the potential might have upon those conclusions. In order to do so, we will use the effective potential
formalism to undertake the computation of the one-loop potential in a simplified theory without fermions or gauge bosons.


\section{The vacuum structure of the Inert model}

The most general 2HDM potential has 14 real parameters, although these may be reduced to 11 using
the reparametrization invariance of the model. In order to avoid tree-level flavour changing neutral currents
--- which are very strongly constrained by experimental measurements --- Glashow, Weinberg and
Paschos~\cite{Glashow:1976nt,Paschos:1976ay} proposed the imposition of a discrete
$Z_2$ symmetry on the lagrangian, so that $\Phi_1\rightarrow \Phi_1$ and
$\Phi_2\rightarrow -\Phi_2$. The resulting scalar potential has but
seven independent real parameters and is written as
\be
V \,=\,
m_{11}^2 |\Phi_1|^2
+ m_{22}^2 |\Phi_2|^2
+ \frac{1}{2} \lambda_1 |\Phi_1|^4
+ \frac{1}{2} \lambda_2 |\Phi_2|^4
+ \lambda_3 |\Phi_1|^2 |\Phi_2|^2
+ \lambda_4 |\Phi_1^\dagger\Phi_2|^2
+
\frac{1}{2} \lambda_5 \left[\left( \Phi_1^\dagger\Phi_2 \right)^2
+ h.c. \right].
\label{eq:pot}
\ee
So that the potential is bounded from below --- thus guaranteed to possess a stable minimum --- the quartic
couplings must obey~\cite{Ma:1978}
\bea
\lambda_1 > 0 & , &  \lambda_2 > 0 \; ,\nonumber \\
\lambda_3 > -\sqrt{\lambda_1 \lambda_2} & , &
\lambda_3 + \lambda_4 - |\lambda_5| > -\sqrt{\lambda_1 \lambda_2} \;.
\label{eq:bfb}
\eea
The $Z_2$ symmetry must be applied to the whole 2HDM lagrangian, otherwise the model would not be renormalisable.
Typically one chooses the $Z_2$ ``charges" of the fermions in such a way that, for instance, only $\Phi_2$ couples to all
fermions (model type I); or such that $\Phi_2$ couples to up-type quarks, $\Phi_1$ to the remaining
fermions~\footnote{In other possibilities - models type X and Y - one varies the coupling of the doublets to the
leptons.}. In the IDM, traditionally, the doublet which is made to couple to all fermions is $\Phi_1$, and
$\Phi_2$ has no coupling to fermions at all.

Let us now consider the possibility of spontaneous symmetry breaking in the potential of Eq.~\ref{eq:pot}.
The doublets may develop neutral vacuum expectation values (vevs), such that $\langle\Phi_1 \rangle = ( 0 \, , \, v_1)^T/\sqrt{2}$ and $\langle\Phi_2 \rangle = ( 0 \, , \, v_2)^T/\sqrt{2}$. Depending on the values of the
parameters of the potential, there are three possible neutral extrema~\cite{Ma:1978}:
\begin{itemize}
\item Both $v_1$ and $v_2$ are non-zero and the vacuum breaks the $Z_2$ symmetry. This is the ``normal" 2HDM
vacuum, interesting in its own right, but not what we wish to consider here.
\item The vev $v_1$ is non-zero, but $v_2 = 0$. This is the {\em inert vacuum}: the $Z_2$ symmetry is preserved,
the fermions acquire a mass through $v_1$ and the Higgs-like scalar $h$ is the real neutral component of $\Phi_1$. The
remaining components of $\Phi_1$ correspond to the Goldstone bosons $G$ and $G^\pm$, and the
doublet $\Phi_2$'s components originate the rest of the scalars --- $H$, $A$ and $H^\pm$, none of which couple to fermions.
The lightest of these is the dark matter candidate~\footnote{Usually taken to be neutral, though choices of parameters
are possible such that the charged scalar is the lightest state.}.
\item The vev $v_2$ is non-zero, but $v_1 = 0$. This is the {\em pseudo-inert vacuum}, or {\em inert-like vacuum}:
a $Z_2$ symmetry is also preserved, but since the fermions only couple to $\Phi_1$ they are massless in this
vacuum. Thus this vacuum is not an acceptable description of reality, and should be avoided as unphysical.
\end{itemize}
Trivial calculations allow us to establish that, for the inert vacuum, one has
\be
v_1^2 \,=\,-\,\frac{2\,m_{11}^2}{\lambda_1},\;\;\;\mbox{provided}\;\;\;m_{11}^2<0.
\ee
Likewise, for the pseudo-inert vacuum, one must have
\be
v_2^2 \,=\,-\,\frac{2\,m_{22}^2}{\lambda_2},\;\;\;\mbox{provided}\;\;\;m_{22}^2<0.
\ee
It has been shown~\cite{Ivanov:2006yq,Ivanov:2007de,Barroso:2007rr} that minima which break different symmetries cannot
coexist. As such, if the potential has parameters such that a minimum with non-zero $(v_1\,,\,v_2)$
exists, then no inert or pseudo-inert minimum exists. However, inert and pseudo-inert vacua can and
do coexist in the model, provided the following conditions are met, at tree-level:
\be
 \mbox{{\em Inert and pseudo-inert minima can coexist in the potential if}} \;\;m_{11}^2<0 \;\;\mbox{{\em and}}\;\; m_{22}^2<0.
\label{eq:sta}
\ee
It is simple to show that there exists an analytical relation between the values of the potential
at the inert minimum ($V_I$) and the pseudo-inert one ($V_{PI}$). Let $v_1 = v$ be the vev value
at the inert minimum, and $v_2 = v^\prime$ the vev at the coexisting pseudo-inert vacuum. Then,
we have
\bea
V_I \,-\,V_{PI} &=& \frac{1}{2} \left( \frac{m^4_{22}}{\lambda_2} \,-\,
\frac{m^4_{11}}{\lambda_1}\right)
\label{eq:difV01} \\
 & & \nonumber \\
 & = & \frac{1}{4} \left[ \left( \frac{m^2_{H^\pm}}{{v^\prime}^2}\right)_{PI} \,-\,
   \left(\frac{m^2_{H^\pm}}{v^2}\right)_{I}\right]\,v^2\,{v^\prime}^2 \;\;\;
\label{eq:difV02}
\eea
where in the second formula we use the value of the squared charged masses at each minimum: in the
inert one, we have
\be
\left(m^2_{H^\pm}\right)_I = m_{22}^2 + \frac{1}{2}\,\lambda_3 \,v^2\;.
\ee
An analogous expression,
with the exchanges $m_{22} \leftrightarrow m_{11}$ and $v \leftrightarrow v^\prime$, is
valid for the charged mass at the pseudo-inert vacuum.

These expressions relating the relative depths of the potential at each of the coexisting
minima are obtained at tree-level. They show that none of the two minima is preferred to the other -
depending on the model's parameters, either minima can be the global one of the theory.
In the rest of this work, we wish to analyse how they
change when the one-loop contributions to the potential are taken into account. In particular,
we wish to investigate whether an ``inversion" of the inert and pseudo-inert minima depths can
occur once the one-loop contributions are considered.

\section{The one-loop 2HDM potential}

At one-loop, the effective potential is given by (in the $\overline{MS}$ scheme, in the Landau gauge)
\be
V\,=\,V_0 \,+\, V_1\,,
\label{eq:pott}
\ee
with $V_0$ given by Eq.~\ref{eq:pot} and the one-loop contribution equal to
\be
V_1 = \frac{1}{64\pi^2}\, \sum_\alpha n_\alpha \,m^4_\alpha(\varphi_i) \left[\log\left(\frac{m^2_\alpha(\varphi_i)}{\mu^2}\right)\,-\,\frac{3}{2}\right]
\label{eq:pot1}
\ee
where $\mu$ is the renormalization scale chosen and the $m_\alpha(\varphi_i)$ are the
field-dependent mass eigenvalues of all particles present in the theory.
In the following analysis we have fixed the value of $\mu$ to be 200 GeV, a mass scale
close enough to the mass scales of the scalars we will be considering. This should be
enough for our purposes --- let us recall that the effective potential of Eq.~\ref{eq:pott}
is renormalization scale independent up to two-loop effects and a good perturbative approximation
provided $\mu$ is of the order of the mass scales involved (thus rendering the logarithms in
Eq.~\ref{eq:pot1} ``small").

In all that follows, we will consider a 2HDM without fermions or gauge bosons.
In other words, a model with two hypercharge 1 doublets, with a global $SU(2)_W\times U(1)_Y$ symmetry. This
toy model, as we will show,  will allow us to ascertain the main features of the one-loop contributions
which interest us. In future works a realistic model will be considered.
The sum over $\alpha$ in Eq.~\ref{eq:pot1} runs therefore from 1 to 8 (all the scalar eigenstates,
though some of them are degenerate, such as the two charged goldstones and charged scalars).
The $\varphi_i$ are the eight real components of the doublets.
The factor $n_\alpha$ counts the number of degrees of freedom corresponding to each particle,
and is in general given, for a particle of spin $s_\alpha$, by
\be
n_\alpha \,=\, (-1)^{2s_\alpha}\,Q_\alpha C_\alpha (2 s_\alpha + 1),
\ee
where $Q_\alpha$ is 1 for uncharged particles and 2 for charged ones; $C_\alpha$ counts the number of
colour degrees of freedom (for particles without colour it equals 1, for particles with colour, 3).
Then, the first derivatives of the one-loop potential are given by (dropping the
explicit field dependence in the masses for simplicity of notation)
\be
\frac{\partial V}{\partial \varphi_i} = \frac{\partial V_0}{\partial \varphi_i} \,+\,
\frac{1}{32\pi^2}\, \sum_\alpha m^2_\alpha \,\frac{\partial m^2_\alpha}{\partial \varphi_i} \left[\log\left(\frac{m^2_\alpha}{\mu^2}\right)\,-\,1\right].
\label{eq:dV}
\ee

Equation~\eqref{eq:dV} can be considerably simplified for the computation of the inert
(and pseudo-inert) vacuum. In the inert case, we have $\langle r_1\rangle = v_1/\sqrt{2}$,
and all remaining $\varphi_i = 0$. An explicit calculation has shown that all derivatives
of the one-loop potential with respect to the fields $\varphi_i$, except
the one with respect to the real neutral component of $\Phi_1$,
are  trivially equal to zero for the inert minimum. Due to the conventions we have chosen,
performing this derivative is equivalent to
differentiating with respect to $v_1$. Thus we obtain~\footnote{Notice the factor of ``2"
affecting both charged scalar contributions in Eq.~\ref{eq:dv1},
counting the number of states effectively present.}
\bea
\frac{1}{v_1}\frac{\partial V}{\partial v_1} & = &  m_{11}^2 + \frac{1}{2}\,\lambda_1\,v_1^2 \,+
\nonumber \\
 & & \frac{1}{32\pi^2} \,\left\{
 \lambda_1\,m^2_{G_0}\,\left[\log\left(\frac{m^2_{G_0}}{\mu^2}\right)\,-\,1\right] \,+\,  3\,\lambda_1\,m^2_{h_0}\,\left[\log\left(\frac{m^2_{h_0}}{\mu^2}\right)\,-\,1\right] \,+ \right.
 \nonumber \\
 & &  \lambda_{345}
 \,m^2_{H_0}\,\left[\log\left(\frac{m^2_{H_0}}{\mu^2}\right)\,-\,1\right] \,+\, \bar{\lambda}_{345}
 \,m^2_{A_0}\,\left[\log\left(\frac{m^2_{A_0}}{\mu^2}\right)\,-\,1\right] \,+
 \nonumber \\
 & & \left.  2\,\lambda_3\,m^2_{H^\pm_0}\,\left[\log\left(\frac{m^2_{H^\pm_0}}{\mu^2}\right)\,-\,1\right] \,+\,
 2\,\lambda_1\,m^2_{G^\pm_0}\,\left[\log\left(\frac{m^2_{G^\pm_0}}{\mu^2}\right)\,-\,1\right]
\right\} \,=\, 0,
\label{eq:dv1}
\eea
where the tree-level scalar masses at the inert minimum are given by
\begin{align}
m^2_{G_0} = m_{11}^2 + \frac{1}{2}\,\lambda_1\,v_1^2,\;\;\;
& \;\;\;
m^2_{G^\pm_0} = m_{11}^2 + \frac{1}{2}\,\lambda_1\,v_1^2,
\label{eq:mGch0} \\
m^2_{h_0} = m_{11}^2 + \frac{3}{2}\,\lambda_1\,v_1^2,
\;\;\; & \;\;\;
m^2_{H_0} = m_{22}^2 + \frac{1}{2}\,\lambda_{345}\,v_1^2,
\label{eq:mH0}\\
m^2_{A_0} = m_{22}^2 + \frac{1}{2}\,\bar{\lambda}_{345}\,v_1^2,
\;\;\; & \;\;\;
m^2_{H^\pm_0} = m_{22}^2 + \frac{1}{2}\,\lambda_3 \,v_1^2,
\label{eq:mHch0}
\end{align}
At the tree-level inert minimum, these tree-level Goldstone masses would be identically
zero. At the one-loop minimum, however, that is no longer so --- one must compute the
full one-loop expressions for $m^2_{G}$ and $m^2_{G^\pm}$ and verify that they are zero
using the full one-loop minimization conditions. We have performed that check and are thus
confident of our one-loop minimization procedure.

In order to ensure we are at a one-loop minimum, all one-loop scalar masses had to be computed.
We worked under the effective potential approximation, assuming that the squared scalar masses
are given by the second derivatives of the one-loop potential. This has been
proven~\cite{Ellis:1990nz,Ellis:1991zd,Brignole:1991pq}
to be a very good approximation to the true one-loop masses. The calculation is made more
difficult by the need to maintain the field dependency in the eigenvalues $m_\alpha(\varphi_i)$
when performing the derivatives.

\section{One-loop inert and pseudo-inert minima}

\begin{figure}[ht]
\centering
\includegraphics[height=8cm,angle=0]{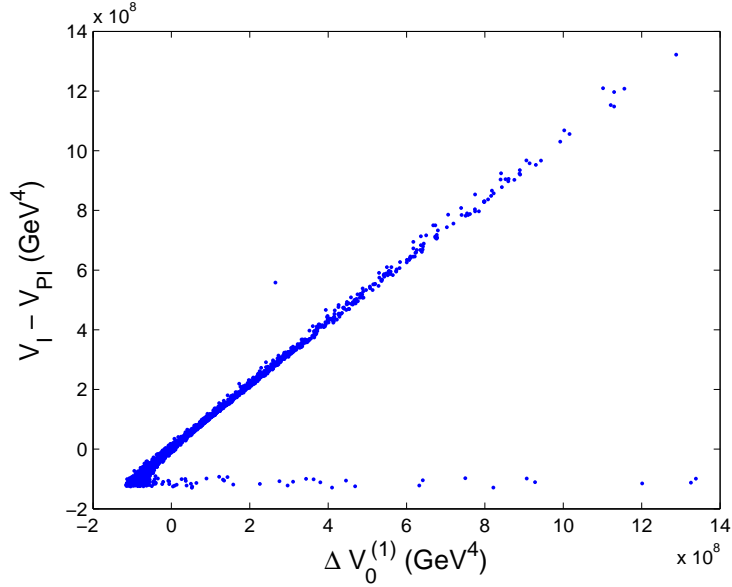}
\caption{One-loop computed difference in inert and pseudo-inert minima depths ($V_I \,-\,V_{PI}$) {\em versus} the
tree-level expected depth difference given by Eq.~\ref{eq:difV01}.}
\label{fig:dV1}
\end{figure}

The one-loop contributions to inert vacua have been studied in great detail (using a different
renormalization approach) in~\cite{Gil:2012}. However, in that work the main topic of analysis was
$T\neq 0$ contributions to the effective potential and its phenomenology. Here we are interested
in the possibility of minima inversion going from the tree-level to the one-loop potential.
Our procedure consisted in scanning the 2HDM parameter space, computing the one-loop
effective potential and its (first and second) derivatives, such that:
\begin{itemize}
\item All tree-level bounded from below conditions were obeyed. This should, to first
approximation, ensure the one-loop potential is also limited from below.
\item Simultaneous inert and pseudo-inert vacua coexist for the choice of parameters made.
The inert vacuum is simply defined as the one with $v_1 = 246$ GeV and $v_2 = 0$, the
pseudo-inert vacuum is such that $v_1 = 0$, $v_2 \neq 0$.
\item We then computed all squared scalar masses at both vacua and demanded they are all
positive (minus the Goldstone boson masses at both minima, which we verified are equal to zero).
We are then assured that we have coexisting minima.
\end{itemize}
With a scan of over 4000 points in 2HDM parameter space with coexisting minima, the
comparison of the one-loop potential inert and pseudo-inert minima depths is shown
in figure~\ref{fig:dV1}. In this plot, we show the difference in value of the inert and pseudo-inert
one-loop minima ($V_I \,-\,V_{PI}$) against the tree-level expected difference in depths from
Eq.~\ref{eq:difV01}, {\em i.e.}
\be
\Delta V_0^{(1)} \,=\, \frac{1}{2} \left( \frac{m^4_{22}}{\lambda_2} \,-\,
\frac{m^4_{11}}{\lambda_1}\right).
\ee
The conclusions to draw from figure~\ref{fig:dV1} are several:
\begin{itemize}
\item The tree-level formula from Eq.~\ref{eq:difV01} is a very good approximation to the one-loop
potential depth difference.
\item It is not however {\em perfect}, since there are clearly deviations from it in the one-loop
results.
\item Furthermore, one finds points for which the inert and pseudo-inert minima are {\em inverted}
going from tree-level to one-loop: if at tree-level the inert minimum was expected to be the deepest,
that is no longer so at one-loop.
\item Confirming that perturbation theory still holds, such inversions are {\em rare} (they have only
occurred for about 3\% of all simulated points) and only occur when both minima are close to degenerate.
\end{itemize}
An interesting observation, though, is obtained if one performs the comparison between $V_I \,-\,V_{PI}$
and the tree-level depth difference formula provided by Eq.~\ref{eq:difV02}, {\em i.e.}, we are now
comparing with
\be
\Delta V_0^{(2)} \,=\, \frac{1}{4} \left[ \left( \frac{m^2_{H^\pm}}{{v^\prime}^2}\right)_{PI} \,-\,
   \left(\frac{m^2_{H^\pm}}{v^2}\right)_{I}\right]\,v^2\,{v^\prime}^2.
   \label{eq:del2}
\ee
We use this formula with the values of the one-loop charged masses. As one sees in
\begin{figure}[ht]
\centering
\includegraphics[height=8cm,angle=0]{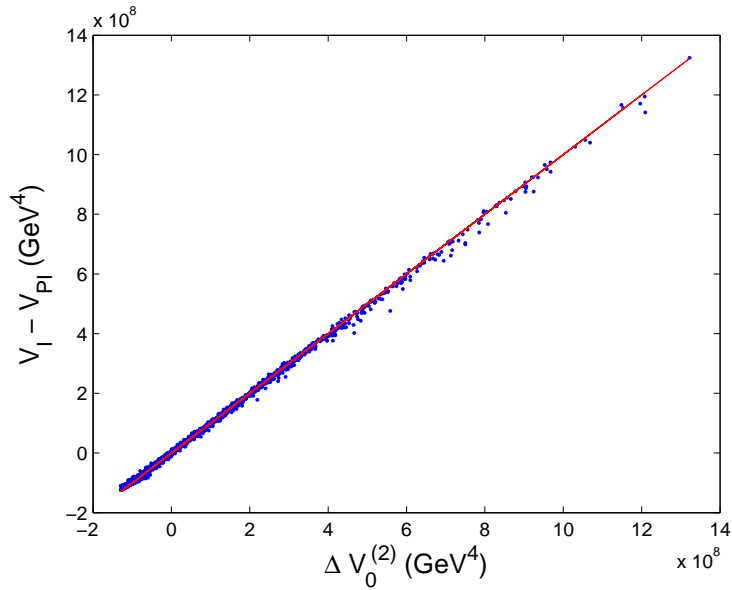}
\caption{One-loop computed difference in inert and pseudo-inert minima depths ($V_I \,-\,V_{PI}$) {\em versus} the
tree-level expected depth difference now given by Eq.~\ref{eq:difV02}.}
\label{fig:dV2}
\end{figure}
figure~\ref{fig:dV2}, the one-loop difference in potential depths is almost perfectly reproduced
by Eq.~\ref{eq:del2} --- the red line in the plot would correspond to the perfect equality
$V_I \,-\,V_{PI} \,=\,\Delta V_0^{(2)}$, and we see there are very little deviations from it.
And in fact, the inversions in depth between both minima now only occur for 0.5\% of the points
simulated (once again, and reassuringly, only for nearly-degenerate potentials).

\begin{figure}[ht]
\centering
\includegraphics[height=8cm,angle=0]{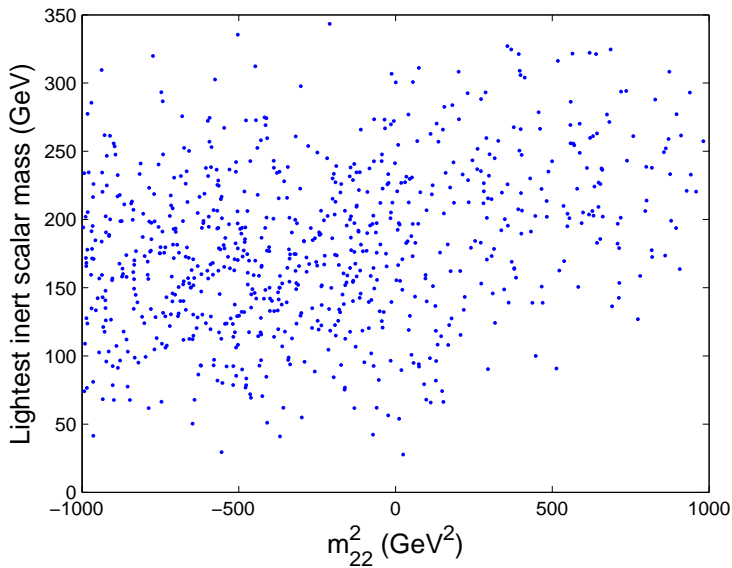}
\caption{Lightest inert scalar mass at the inert minimum {\em versus} the $m_{22}^2$ quadratic parameter.}
\label{fig:m22}
\end{figure}

Equally interesting conclusions are drawn from figure~\ref{fig:m22}, where we plot the mass of the
lightest inert scalar at the inert vacuum (meaning, the dark matter candidate) against the quadratic
parameter $m_{22}^2$, for all points scanned with coexisting inert and pseudo-inert minima. Though the
range of masses for the dark matter candidate is interesting in its own merit, it's the horizontal
axis that provides us with a very interesting fact: {\em at one-loop, we can have coexistence of
inert and pseudo-inert minima even if $m_{22}^2 >0$}. This, according to the statement presented
in~\ref{eq:sta}, was not possible at tree-level!

\section{Conclusions}

The one-loop analysis of inert and pseudo-inert coexisting minima has shown that an inversion of
the relative depths of these
minima can be caused by radiative corrections. The results obtained indicate that this possibility can
occur when the tree-level minima are nearly degenerate, so that this result does not put into
question the validity of the perturbative approximation. We have further shown that the region of
2HDM parameter space where one can expect coexistence of inert and pseudo-inert minima is
{\em extended} at the one-loop level, compared to tree-level expectations.

The formulae deduced at tree-level for the depth difference at both minima do not remain valid
at the one-loop level. However, expressing the difference in potential depths in terms of physical
charged masses almost perfectly reproduces the one-loop results. This almost suggests that the
tree-level formula of Eq.~\ref{eq:difV02} might indeed end up valid at the one loop level --- remember
that we computed all one-loop masses within the effective potential approximation, so it is within
the realm of possibilities that a completely accurate mass calculation may lead to full agreement
between Eq.~\ref{eq:difV02} and the one-loop results. That would suggest that the tree-level
deduction of that equation had somehow ``stumbled" upon an exact formula.

Finally, all of these conclusions were drawn in the context of a toy model devoid of gauge bosons and
fermions. But the conclusions are interesting by themselves, since they show that within the scalar sector
alone one may already expect deviations from the tree-level formulae deduced for the potential
depths' difference. Certainly it is to be expected that the inclusion of fermions will lead to great
differences in tree-level and one-loop comparisons between inert and pseudo-inert minima --- the fact that
in the pseudo-inert minima all fermions are massless, as opposed to what happens for the inert minimum, leads
one to expect further inversions of the minima, for instance. But the fact remains that the analysis
herein detailed already shows that such inversions are not caused by the fermionic or gauge sector of
the theory, they're rather already present within the scalar sector at the one-loop level.

\begin{acknowledgments}
The work of P.M.F. is supported in part by the Portuguese
\textit{Funda\c{c}\~{a}o para a Ci\^{e}ncia e a Tecnologia} (FCT)
under contract PTDC/FIS/117951/2010, by FP7 Reintegration Grant, number PERG08-GA-2010-277025,
and by PEst-OE/FIS/UI0618/2011. The work of B.S. is supported by the Polish National Science Centre grant
PRELUDIUM under the decision number DEC-2013/11/N/ST2/04214. P.M.F. would also like to thank the support and hospitality
of the University of Toyama during the HPNP-2015 Workshop.
\end{acknowledgments}

\bigskip 

\begin{thebibliography}{99} 
\bibitem{Lee:1973iz}
 T.~D.~Lee,
 Phys.\ Rev.\  D {\bf 8} (1973) 1226.

\bibitem{Branco:2011iw}
  G.~C.~Branco, P.~M.~Ferreira, L.~Lavoura, M.~N.~Rebelo, M.~Sher and J.~P.~Silva,
  Phys.\ Rept.\  {\bf 516}, 1 (2012)
  [arXiv:1106.0034 [hep-ph]].

\bibitem{:2012gk}
G.~Aad {\it et al.}  [ATLAS Collaboration],
Phys.\ Lett.\ B \textbf{716}, 1 (2012)
[arXiv:1207.7214 [hep-ex]].

\bibitem{:2012gu}
S.~Chatrchyan \textit{et al.}  [CMS Collaboration],
Phys.\ Lett.\ B \textbf{716}, 30 (2012)
[arXiv:1207.7235 [hep-ex]].

\bibitem{Chen:2013kt}
 C.~-Y.~Chen and S.~Dawson,
 arXiv:1301.0309 [hep-ph].

\bibitem{Belanger:2012gc}
 G.~Belanger, B.~Dumont, U.~Ellwanger, J.~F.~Gunion and S.~Kraml,
 arXiv:1212.5244 [hep-ph].

\bibitem{Chang:2012ve}
 S.~Chang, S.~K.~Kang, J.~-P.~Lee, K.~Y.~Lee, S.~C.~Park and J.~Song,
with mass around 125 GeV,''
 arXiv:1210.3439 [hep-ph].

\bibitem{Ferreira:2011aa}
 P.~M.~Ferreira, R.~Santos, M.~Sher and J.~P.~Silva,
 Phys.\ Rev.\ D {\bf 85}, 077703 (2012)
 [arXiv:1112.3277 [hep-ph]].

 \bibitem{Chowdhury:2015yja}
  D.~Chowdhury and O.~Eberhardt,
  arXiv:1503.08216 [hep-ph].

\bibitem{Ma:1978}
N.~G. Deshpande and E.~Ma,
\newblock Phys.Rev. {\bf D18}, 2574 (1978).

\bibitem{Cao:2007}
Q.-H. Cao, E.~Ma, and G.~Rajasekaran,
\newblock Phys.Rev. {\bf D76}, 095011 (2007), arXiv:0708.2939.

\bibitem{Barbieri:2006}
R.~Barbieri, L.~J. Hall, and V.~S. Rychkov,
\newblock Phys.Rev. {\bf D74}, 015007 (2006), arXiv:hep-ph/0603188.

\bibitem{LopezHonorez:2006}
L.~Lopez~Honorez, E.~Nezri, J.~F. Oliver, and M.~H. Tytgat,
\newblock JCAP {\bf 0702}, 028 (2007), arXiv:hep-ph/0612275.

\bibitem{Dolle:2009}
E.~M. Dolle and S.~Su,
\newblock Phys.Rev. {\bf D80}, 055012 (2009), arXiv:0906.1609.

\bibitem{Honorez:2010}
L.~Lopez~Honorez and C.~E. Yaguna,
\newblock JHEP {\bf 1009}, 046 (2010), arXiv:1003.3125.

\bibitem{Sokolowska:2011}
D.~Soko{\l}owska,
\newblock (2011), arXiv:1107.1991.

\bibitem{Gustafsson:2012}
M.~Gustafsson, S.~Rydbeck, L.~Lopez-Honorez, and E.~Lundstrom,
\newblock (2012), arXiv:1206.6316.

\bibitem{Swiezewska:2012}
B.~{\'S}wie{\.z}ewska and M.~Krawczyk,
\newblock Phys.Rev. {\bf D88}, 035019 (2013), arXiv:1212.4100.

\bibitem{Krawczyk:2013jhep}
M.~Krawczyk, D.~Soko{\l}owska, P.~Swaczyna, and B.~{\'S}wie{\.z}ewska,
\newblock JHEP {\bf 1309}, 055 (2013), arXiv:1305.6266.

\bibitem{Arhrib:2013}
A.~Arhrib, Y.-L.~S. Tsai, Q.~Yuan, and T.-C. Yuan,
\newblock JCAP {\bf 1406}, 030 (2014), arXiv:1310.0358.

\bibitem{Goudelis:2013}
A.~Goudelis, B.~Herrmann, and O.~Stål,
\newblock JHEP {\bf 1309}, 106 (2013), arXiv:1303.3010.

\bibitem{Chakrabarty:2015}
N.~Chakrabarty, D.~K. Ghosh, B.~Mukhopadhyaya, and I.~Saha,
\newblock (2015), arXiv:1501.03700.

\bibitem{Ginzburg:2010}
I.~Ginzburg, K.~Kanishev, M.~Krawczyk, and D.~Sokolowska,
\newblock Phys.Rev. {\bf D82}, 123533 (2010), arXiv:1009.4593.

\bibitem{Sokolowska:2011t}
D.~Sokolowska,
\newblock (2011), arXiv:1104.3326.

\bibitem{Kanemura:2004}
S.~Kanemura, Y.~Okada, and E.~Senaha,
\newblock Phys.Lett. {\bf B606}, 361 (2005), arXiv:hep-ph/0411354.

\bibitem{Gil:2012}
G.~Gil, P.~Chankowski, and M.~Krawczyk,
\newblock Phys.Lett. {\bf B717}, 396 (2012), arXiv:1207.0084.

\bibitem{Cline:2013}
J.~M. Cline and K.~Kainulainen,
\newblock Phys.Rev. {\bf D87}, 071701 (2013), arXiv:1302.2614.

\bibitem{Chowdhury:2011}
T.~A. Chowdhury, M.~Nemevsek, G.~Senjanovic, and Y.~Zhang,
\newblock JCAP {\bf 1202}, 029 (2012), arXiv:1110.5334.

\bibitem{Ma:2006neutrino}
E.~Ma,
\newblock Phys.Rev. {\bf D73}, 077301 (2006), arXiv:hep-ph/0601225.

\bibitem{Gustafsson:2012neutrino}
M.~Gustafsson, J.~M. No, and M.~A. Rivera,
\newblock Phys.Rev.Lett. {\bf 110}, 211802 (2013), arXiv:1212.4806,
\newblock Erratum-ibid. 112 (2014) 25, 259902.

\bibitem{Davoudiasl:2014pya}
  H.~Davoudiasl and I.~M.~Lewis,
  Phys.\ Rev.\ D {\bf 90} (2014) 3,  033003.

 \bibitem{Glashow:1976nt}
 S.~L.~Glashow and S.~Weinberg,
 Phys.\ Rev.\  D {\bf 15} (1977) 1958.

\bibitem{Paschos:1976ay}
 E.~A.~Paschos,
 Phys.\ Rev.\  D {\bf 15} (1977) 1966.

\bibitem{Barroso:2007rr}
  A.~Barroso, P.~M.~Ferreira and R.~Santos,
  Phys.\ Lett.\ B {\bf 652} (2007) 181.

\bibitem{Ivanov:2006yq}
 I.~P.~Ivanov,
 Phys.\ Rev.\  D {\bf 75} (2007) 035001
 [Erratum-ibid.\  D {\bf 76} (2007) 039902]
 [arXiv:hep-ph/0609018].

\bibitem{Ivanov:2007de}
 I.~P.~Ivanov,
 Phys.\ Rev.\  D {\bf 77},  015017 (2008).

\bibitem{Ellis:1990nz}
  J.~R.~Ellis, G.~Ridolfi and F.~Zwirner,
  Phys.\ Lett.\ B {\bf 257} (1991) 83.

\bibitem{Ellis:1991zd}
  J.~R.~Ellis, G.~Ridolfi and F.~Zwirner,
  Phys.\ Lett.\ B {\bf 262} (1991) 477.

\bibitem{Brignole:1991pq}
  A.~Brignole, J.~R.~Ellis, G.~Ridolfi and F.~Zwirner,
  Phys.\ Lett.\ B {\bf 271} (1991) 123.


\end{thebibliography}

\end{document}